\documentclass[
 manuscript,
 ]{aastex631}
\usepackage{amsmath}

\submitjournal{ApJ}

\shorttitle{Knudsen number as a skewness parameter}
\shortauthors{Gallo-M\'endez, Viñas, \& Moya}
\graphicspath{{./}{figures/}}

\begin{document}

\title{Knudsen number as a non-thermal parameter: possible origin of skewness in space plasma distributions}

\author[0000-0003-4988-4348]{Iván Gallo-Méndez}
\email{ivan.gallo@ug.uchile.cl}
\affiliation{Departamento de F\'isica, Facultad de Ciencias, Universidad de Chile, Santiago, Chile.}

\author{Adolfo F. Viñas}
\email{afvinas@gmail.com}
\affiliation{Department of Physics \& the Institute for Astrophysics and Computational Sciences (IACS), Catholic University of America, Washington-DC, 20064, US}
\affiliation{NASA Goddard Space Flight Center, Emeritus Scientist, Heliospheric Science Division, Mail Code 673, Greenbelt MD 20771, USA.}

\author[0000-0002-9161-0888]{Pablo S. Moya}
\email{pablo.moya@uchile.cl}
\affiliation{Departamento de F\'isica, Facultad de Ciencias, Universidad de Chile, Santiago, Chile.}



\begin{abstract}

Non-Maxwellian distributions and their origins in space plasma have attracted significant attention due to their prevalence and impact on various astrophysical and space-related phenomena. This paper presents a theoretical study of the consequences of incorporating a Skew-Kappa distribution to describe the non-thermal electron distribution in the solar wind. By introducing a Krook-like term into the Boltzmann equation to represent collision effects, we investigate the dependence of the skewness parameter on plasma macro-dynamics. Our analysis focuses on understanding the departure from thermal equilibrium and the statistical behavior of the plasma under the influence of collisional processes. By analyzing the Boltzmann transport equation adapted to space plasma, we derive expressions for the skewness parameter as a function of plasma parameters and the collision effect. Our results provide valuable information on the relationship between skewness, collisional dynamics, and the statistical properties of space plasmas, namely $\delta \sim K_N$, the relationship between the skewness parameter and the effective Knudsen number. This study contributes to a deeper understanding of non-Maxwellian distributions and their role in astrophysical and space plasma phenomena.
\end{abstract}



\section{Introduction} 
\label{sec:intro}
Non-thermal features in the Velocity Distribution Function (VDF) have been the subject of significant interest in plasma physics, particularly in space plasmas. An example is the electron population in the Solar Wind (SW) plasma, where measurements of the velocity distribution function have revealed high-energy tails that lie outside the scope of traditional Boltzmann-Gibbs statistics. Hence, these are a class of non-Maxwellian or non-exponential distributions. Examples of such functions include Kappa distributions in several contexts of interplanetary plasma \citep{vasyliunas1968survey}, regularized Kappa distributions in astrophysical and space plasmas \citep{scherer2018regularized, husidic2020linear}, $q$-distributions \citep{tsallis1988possible} studied within the framework of non-extensive statistics in complex systems, among other distributions beyond the typical Maxwellian equilibrium. For instance, plasmas in the planetary magnetosphere deviate significantly from Maxwellian distributions due to the presence of high-energy particles. Measurements of plasma velocity distributions in the solar wind, planetary magnetosphere, and magnetosheaths have revealed the prevalence of non-Maxwellian distributions, particularly in the electron population, as well as in the distribution of electromagnetic fields \citep{nieves2008solar, bruno2013solar}. In many cases, these distributions exhibit a supra-thermal power-law tail at high energies, which has been successfully modeled by the well-known Kappa-like distributions \citep{pierrard2010kappa, lazar2015destabilizing, lazar2021kappa}, namely:

\begin{equation}
    f_\kappa(\vec{v}) = n_0 B_\kappa \left[1 + \frac{1}{\kappa - \alpha} \frac{v^2}{w^2}\right]^{-(\kappa + 1)} \ ,
\label{KappaDistribution}
\end{equation}
where
\begin{equation}
    B_\kappa = \left[\pi w^2 \left(\kappa - \alpha \right)\right]^{-3/2} \frac{\Gamma(\kappa + 1)}{\Gamma(\kappa - 1/2)}\ .
\label{B_k}
\end{equation}
Here, we used $w$ to refer to thermal velocity, and $\alpha$ to replace two specific values, $0$ and $3/2$, which correspond to the most commonly used kappa distributions. For the first value ($\alpha = 0$), the thermal speed for the kappa distribution, defined as the second moment of the distribution, remains the
same as in the Maxwellian case. It leads to a definition of the kinetic temperature dependent on the kappa parameter. The second value ($\alpha = 3/2$) represents the case where the thermal speed is assumed to be kappa-dependent, and the kinetic temperature remains the same as in the Maxwellian case, which is kappa-independent \citep{daniel2024magnetic}.

Unlike classical systems in thermal equilibrium, where particles exhibit limited or no correlations among their velocities or energies, space plasmas are frequently in stationary states that deviate from thermal equilibrium. The velocity distributions in these plasmas are stabilized into functions that differ from the traditional Maxwell-Boltzmann formulation because these systems are characterized by long-range interactions that induce correlations among particles, resulting in collective behaviors \citep{hasegawa1985plasma, gallo2022langevin, gallo2023understanding}. This is primarily due to the high spatio-temporal correlation between the plasma and self-generated electromagnetic fields \citep{matthaeus2016ensemble, Jana2023Studies}. However, observing these high-energy tails as non-thermal features in the distribution function is not the only possibility. As demonstrated by several authors, it is also feasible to discern frequent cases where the VDF of electrons in the solar wind exhibits anomalies, such as being asymmetric with respect to the mean velocity of the particles in the direction of the background magnetic field \citep{pilipp1987characteristics, nieves2008solar}. 

Observational data have demonstrated a distinctive electron distribution in the solar wind, characterized by a substantial, dense core component and a less dense supra-thermal population. The core component, characterized by thermal energies around 10 eV, accounts for approximately 95 \% of the total electron density. In contrast, the supra-thermal population comprises roughly 5 \% of the electron density and comprises two energetic sub-populations. The first sub-population, known as the ``Halo", exhibits characteristics that can be approximated by a Kappa-like distribution with a higher temperature compared to the core distribution. The second sub-population, the anisotropic component, or ``Strahl'', consists of electrons moving anti-sunward with a narrow pitch-angle distribution. Previous studies as \cite{rosenbauer1977survey, feldman1978characteristic, lin1981energetic, ogilvie2000electrons, gosling2004dispersionless, pagel2007scattering} have highlighted these electron population characteristics.

Furthermore, it has been observed that the proportion of halo electrons increases while the proportion of strahl electrons decreases with increasing distance from the Sun. This phenomenon suggests that the heliospheric electron halo population contains electrons that have been scattered from the strahl component (\cite{maksimovic2005radial, vstverak2009radial}), providing valuable insights into the complex dynamics of the solar wind electron population. These considerations lead us to identify heat transfer mechanisms inherent to the plasma under conditions similar to those of the solar wind. Some authors \citep{salem2003electron, bale2013electron, halekas2020electrons, halekas2021electron, bervcivc2020coronal} have previously reported this in the case of collisions, demonstrating how the heat flux exhibits a significant dependence on the Knudsen number and confirming the validity of Spitzer \& Harm's theory. Hence, we observe a direct relationship between the electron Knudsen number and the heat flux, and consequently with the asymmetry of the electron velocity distribution function. On the other hand, it should be emphasized that weakly collisional or even collisionless mechanisms can occur in the space plasma. Numerous studies have characterized these processes and elucidated the kinetic consequences of electron weak collisionality, including instabilities in the system \citep{lazar2017firehose, lopez2019particle, lopez2020alternative, shaaban2021advanced}. 

Early evidence that asymmetric Kappa-like electron distributions play a role in regulating solar wind stability was reported by \citet{vinas2015electron}, who analyzed the full electron VDF without separating individual populations. Since that work, the skewed Kappa distribution has been considered as a viable model for describing electron velocity distributions. Building on this idea, \citet{Zenteno2021model} later formalized the CS model, describing the electron VDF as a bi-Maxwellian core plus a single Skew--Kappa distribution that simultaneously captures the suprathermal tails through the $\kappa$ parameter and the field-aligned asymmetry through a single skewness parameter $\delta$. Among other properties, this representation reconciles the observed asymmetry and non-thermal features without invoking separate halo and strahl fits in each case. More recently, the CS model has been applied to in-situ eVDFs measured by the WIND spacecraft at $\sim$1 AU, showing that a single $\delta$ robustly characterizes the asymmetry while recovering heat-flux trends, with $\kappa$ largely decoupled from $\delta$ \citep{eyelade2025characterizing}. These observational results directly motivate the transport-based, collisional perspective developed below, in which the emergence and maintenance of asymmetry can be discussed in terms of macroscopic gradients and collisional effects.

Building on these observational characterizations of non-thermal VDF features and their representation using the CS model, here we ask under what collisional transport conditions a persistent, field-aligned skewness can arise and be maintained in the solar wind. To address this question, we adopt a steady-state Skew-Kappa distribution as an ansatz for the electron VDF and analyze the Boltzmann equation augmented with an effective collisional operator on the right-hand side. We examine the consistency and stability of this description and outline the conditions under which an asymmetry parameter can exist, thereby framing testable implications for heat transport without assuming a priori any fixed relation between $\delta$ and $\kappa$.

This paper is organized as follows. Section \ref{sec:model} provides a comprehensive overview of the theoretical framework, including the Boltzmann Transport Equation and the incorporation of a Krook term on its right-hand side, suitable for an electron-proton plasma. Section \ref{sec:AyR} presents the mathematical analysis that derives the expressions for the skewness parameter and discusses the implications of our findings and their relevance to the study of space plasma. Finally, we conclude our work in Section \ref{sec:conclusions}, highlighting the significance of considering skewness and collisional effects in understanding the complex dynamics of space plasmas. Additionally, we present an Appendix \ref{sec:appendix} that provides the mathematical details of the derivation of an analytical expression for the skewness parameter.

\section{Skew Kappa Distributions as a solution of the Boltzmann Equation}
\label{sec:model}
To describe the non-extensive statistics features of solar wind electrons, we model their distribution by the generalized Boltzmann Transport Equation (BTE)
\begin{equation}
    \frac{\partial f_e}{\partial t} + \vec{v} \cdot \frac{\partial f_e}{\partial \vec{r}} + \frac{q_e}{m_e}\left( \vec{E} + \frac{\vec{v}}{c} \times \vec{B} \right) \cdot \frac{\partial f_e}{\partial \vec{v}} = \left( \frac{\partial f_e}{\partial t} \right)_\text{coll} \ .
\label{BoltzmannEq}
\end{equation}
Note that here we use a collisional term on the right-hand side of the equation as the system driver, which breaks out of the typical steady state. Further, we consider $\vec{E} = \delta\vec{E}$ and $\vec{B} = \delta\vec{B} + \vec{B}_0$, where $\delta$ quantities are self-generated electromagnetic fields fluctuations, while $\vec{B}_0$ is the background magnetic field. Some studies suggest that the solar wind can be modeled with Skew-Kappa distributions \citep{Zenteno2021model, zenteno2022role, Zenteno2023interplay}. This is why we will impose this solution as a steady state of the previous equation, namely:
\begin{equation}
    f_e^{\kappa\delta} = n_e A_{\kappa\delta}\left[ 1 + \frac{v_\perp^2}{(\kappa - \alpha)W_{\perp,e}^2} + \frac{ v_\parallel^2}{(\kappa - \alpha)W_{\parallel,e}^2} + \frac{{\delta_e}}{\kappa - \alpha}\left( \frac{v_\parallel }{W_{\parallel,e}} - \frac{v_\parallel^3}{3W_{\parallel,e}^3} \right)  \right]^{-(\kappa + 1)} \ .
    \label{SkewKappa}
\end{equation}
\begin{figure}[ht!]
\centering
    \includegraphics[width=0.8\textwidth]{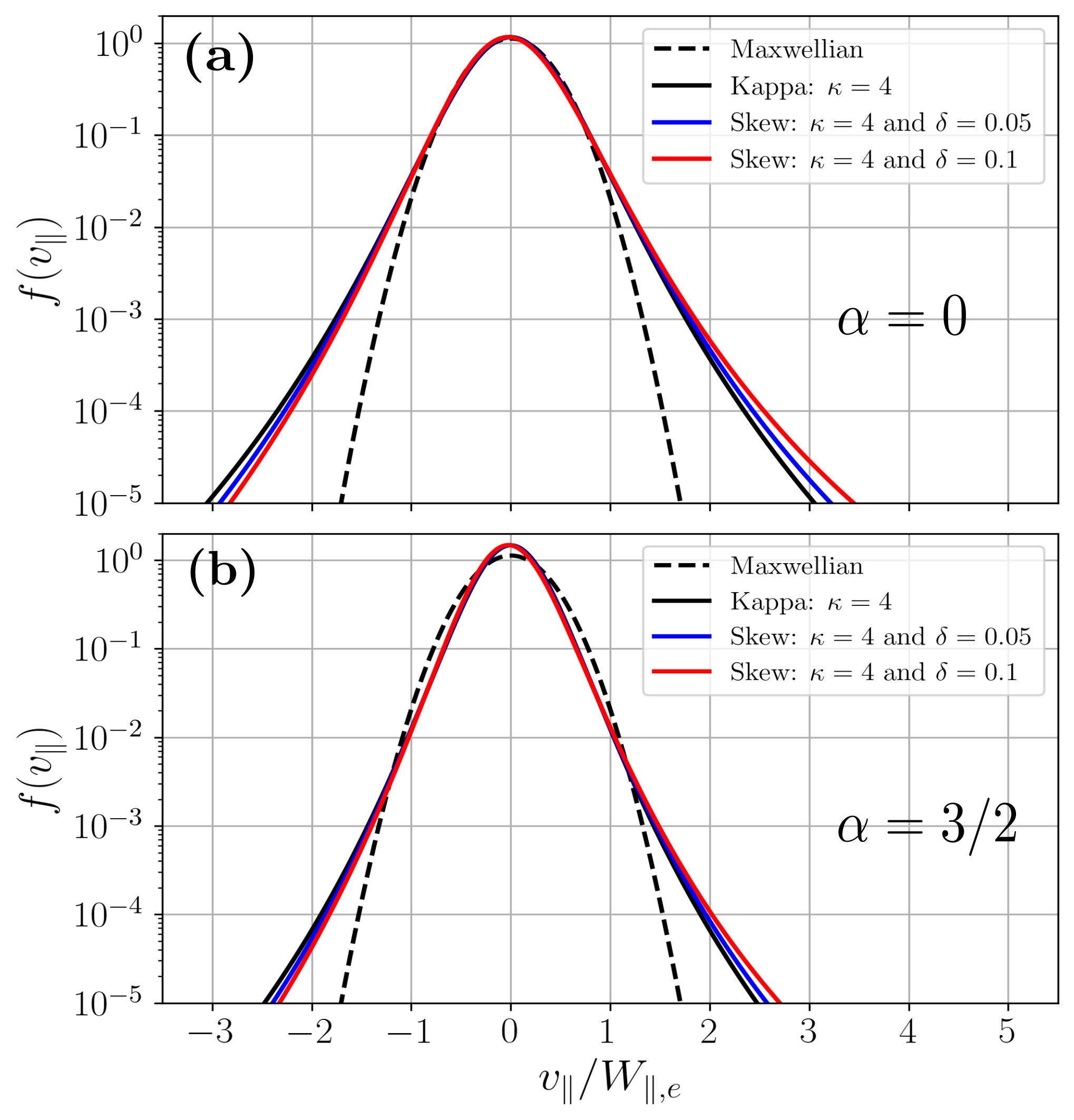}
    \caption{Comparative plot of different types of distribution functions. In the top and bottom figures, the dashed black curves represent a Maxwellian distribution that decays much faster than the other functions. The solid black lines correspond to a Kappa distribution with $\kappa = 4$. In contrast, the blue and red curves represent Skew-Kappa distributions, both with $\kappa = 4$ and a skewness parameter $\delta = 0.05$ and $\delta = 0.1$, respectively. In \textbf{(a)} are plotted  Kappa and Skew-Kappa distributions with $\alpha = 0$. In figure \textbf{(b)} are plotted for $\alpha = 3/2$.}
    \label{comparative_skew}
\end{figure}
Although solar wind electrons comprise thermal and non-thermal electrons (core, halo, and strahl, or core and strahlo) in this study, we focus only on the population of suprathermal particles that deviate from the core and exhibit asymmetry in the distribution function, in addition to high-energy tails. It is also necessary to note that here we consider distributions with small asymmetry, such that the skewness parameter in the Skew-Kappa distribution in \eqref{SkewKappa} satisfies $\delta_e \ll 1$. We present a comparative graph of this function (Figure \ref{comparative_skew}) for different delta values, contrasting it with the Kappa distribution and a typical Maxwellian distribution. In Figures \ref{comparative_skew}(a) and (b), Kappa and Skew-Kappa distributions are plotted with $\alpha = 0$ and $\alpha = 3/2$, respectively. In the top and bottom panels, the dashed black curves represent a Maxwellian distribution that decays much faster than the other functions. The solid black lines correspond to a Kappa distribution with $\kappa = 4$. In contrast, the blue and red curves represent Skew-Kappa distributions, both with $\kappa = 4$ and a skewness parameter $\delta = 0.05$ and $\delta = 0.1$, respectively.

In addition to assuming small asymmetries, we explicitly adopt a separation of timescales between the kinetic evolution and the slow modulation of the skewness parameter. Denoting by $\tau_{\rm kin}$ the characteristic kinetic timescale
(collisional/advective) and by $\tau_{\delta}$ the timescale on which $\delta_e$ evolves, we assume
$\varepsilon \equiv \tau_{\rm kin}/\tau_{\delta} \;\ll\; 1$,
so that, by the chain rule,
\begin{equation}
    \frac{\partial f_e^{\kappa\delta}}{\partial t}
    = \frac{\partial f_e^{\kappa\delta}}{\partial \delta_e}\cdot\frac{\partial \delta_e}{\partial t}
    = \mathcal{O}(\varepsilon) \;\;\;\;\Rightarrow\;\;\;\; \frac{\partial f_e^{\kappa\delta}}{\partial t} \approx 0 \ .
\end{equation}
Operationally, we treat $\delta_e$ as a quasi-constant parameter on $\tau_{\rm kin}$, i.e. $\partial_t \delta_e \approx 0$ to leading order, which justifies taking the Skew–Kappa form in Eq.~ \eqref{SkewKappa} as a stationary ansatz for the Boltzmann Transport Equation, and underpins the perturbative expansion in $\delta_e$ developed below.

Using the small-asymmetry limit observed in the solar wind, we expand to second order in $\delta_e$, perform analytic work on Equation~\eqref{BoltzmannEq}, and thus obtain a dependence of the system's asymmetry parameter on macro-parameters. To do that, we consider a Krook-type collisional operator \citep{bhatnagar1954model, wang2017diffusion, wang2018viscosity, husidic2022toward}, as following,
\begin{equation}
    \left( \frac{\partial f_e^{\kappa\delta}}{\partial t} \right)_\text{coll} \approx - \nu_e(\vec{v}) \left( f^{\kappa\delta}_e - f_e^\kappa \right) \ ,
    \label{Krook}
\end{equation}
This equation preserves the form of the typical Krook approximation for slight deviations from the collisionless steady state. At this point, treating the stationary VDF with collisions is common: perturb the system to first order, obtain the perturbation, and use it to complete the new steady-state solution. In our case, as we set in our model, we impose stationarity by assuming that the system is described by a Skew-Kappa distribution, with a small skewness parameter, thereby satisfying the stated condition.

Note that in \citet{wang2017diffusion, wang2018viscosity}, the authors model the right-hand side of the BTE in a first approximation by treating the collision frequency as constant with respect to particle velocity. However, \cite{husidic2022toward}, in a subsequent approximation, takes into account the expression given by \cite{helander2005collisional} (H\&S) and introduces a velocity-dependent term on the collision frequency for charged particles, given by:
\begin{equation}
    \nu_{ei}(v) = \frac{4\pi n_e z e^4 L^{ei}}{m_e^2 v^3} \ .
    \label{nu_hs}
\end{equation}

To succinctly encapsulate the information conveyed by Equation~\eqref{nu_hs}, we shall rephrase it to elucidate the normalization of the involved variables, thereby providing a redefined formulation:
\begin{equation}
    \nu_{ei}(v) = \frac{4\pi n_e z e^4 L^{ei}}{m_e^2 W^3} \frac{W^3}{v^3} \ .
\end{equation}
In this way, we can discern the dependence on velocity by normalizing the system to a characteristic velocity, $W$, such as the thermal velocity. Namely:
\begin{equation}
    \nu_{ei}(v) = \nu_{e,0} \frac{W^3}{v^3} \ .
    \label{nu_hs_normalized}
\end{equation}
As mentioned, in our scenario, we adopt the Skew-Kappa distribution as the proposed solution, and the implications of this hypothesis will directly affect the collisional operator. Thus, considering the asymmetric and anisotropic Skew-Kappa as a quasi-steady-state solution of the BTE affects the collision frequency, so its functional form differs from Eq.~\eqref{nu_hs_normalized}. To address this issue, we propose a collision frequency that adheres to rational constraints, including convergence to a finite value, denoted by $\nu_{e,0}$, for velocities near zero. Additionally, this frequency exhibits a decay pattern similar to the power-law behavior observed in the H\&S model, as depicted in Eq. \eqref{nu_hs}. 

To achieve this objective, we have adopted the following mathematical expression:
\begin{equation}
    \nu_e(\vec{v}) = \nu_{e,0}\left[ 1 + \frac{v_\perp^2}{(\kappa - \alpha)W_{\perp,e}^2} + \frac{v_\parallel^2}{(\kappa - \alpha)W_{\parallel,e}^2}  \right]^{-3/2} \ .
    \label{nu_model}
\end{equation}
Equation~\eqref{nu_model} represents the most general form of the collision-frequency model adopted in this work, explicitly accounting for the parallel and perpendicular components of the particle velocity. The subsequent restriction to the symmetric and isotropic case, $W_{\perp,e} = W_{\parallel,e} \equiv W$, is introduced solely as a simplifying limit that allows for a more direct comparison with standard results. In particular, in the high-velocity regime, this isotropic limit recovers the well-known scaling $\nu_e \sim v^{-3}$ characteristic of the Helander \& Sigmar. This behavior reflects the underlying assumption of Coulomb collisions dominated by small-angle deflections, for which cumulative collisional effects primarily modify the magnitude of the velocity rather than its direction.

For the purposes of illustration and comparison, we now consider the symmetric and isotropic case in which the parallel and perpendicular thermal velocities are equal. This yields the following outcome:
\begin{equation}
    \nu_e(\vec{v}) = \nu_{e,0}\left[ 1 + \frac{v^2}{(\kappa - \alpha)W^2}\right]^{-3/2} \ .
    \label{isotropic_nu_model}
\end{equation}
\begin{figure}[ht]
\centering
    \includegraphics[width=0.8\textwidth]{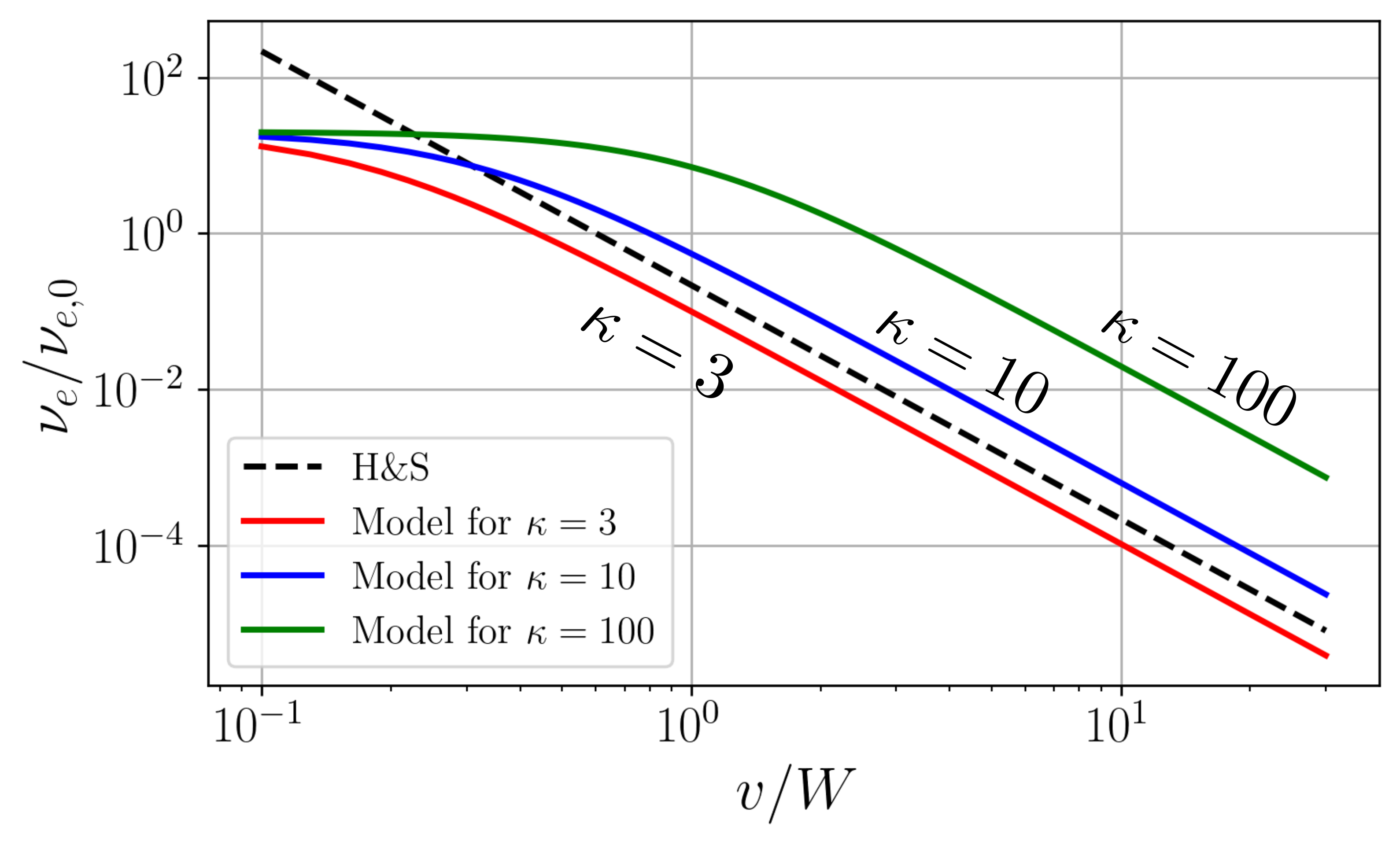}
    \caption{Comparison of collision frequency $\nu_e(v)$ function curves. The dashed black curve represents the collision frequency described by the Helander \& Sigmar model, which exhibits a velocity-dependent collision frequency following a power law. On the other hand, the red curve represents the collision frequency model proposed in Eq. \eqref{isotropic_nu_model} for $\alpha = 0$. In our model, $\nu_e$ converges to a finite value as velocity tends to 0, and also for large values of $\kappa$. At the same time, it decreases according to a power law, at the same rate as H\&S, for sufficiently high velocities.}
    \label{Nu}
\end{figure}

Figure~\ref{Nu} illustrates the behavior of the collision-frequency model in Eq.~\eqref{isotropic_nu_model} for different values of $\kappa$ (with $\alpha=0$), and contrasts it with the Helander \& Sigmar prescription. As shown, the proposed model smoothly interpolates between a nearly constant collision frequency at low velocities and a power-law decay at high velocities, reproducing the expected asymptotic behavior while remaining finite for $v \to 0$. This property makes it particularly suitable for analytical treatments of weakly collisional, quasi-steady electron distributions. Although the general formulation in Eq.~\eqref{nu_model} explicitly distinguishes between parallel and perpendicular velocity components, for the purposes of the present analysis, the corresponding thermal velocities are expected to be comparable. More importantly, the key ingredient controlling the emergence of asymmetry in our framework is the presence of a spatial gradient in the parallel thermal velocity, $W_{\parallel,e}(z)$, rather than the precise degree of temperature anisotropy.

\section{Analysis and Results}
\label{sec:AyR}

Under the assumptions introduced in Section \ref{sec:model}, namely a stationary Skew–Kappa electron distribution and a velocity-dependent Krook-type collision operator, we now proceed to derive an explicit expression for the skewness parameter. Substituting the proposed distribution function, given by Eq. \eqref{SkewKappa}, into the Boltzmann Transport Equation, the kinetic equation reduces to a balance between streaming and collisional relaxation terms. This reduced form constitutes the starting point of our analysis:
\begin{equation}
   v_\parallel \frac{\partial f_e^{\kappa\delta}}{\partial z} = - \nu_e(\vec{v}) \left( f^{\kappa\delta}_e - f_e^\kappa \right) \ .
    \label{reduced_BTE}
\end{equation}
To maintain an intuitive order, we will call Equation \eqref{reduced_BTE} the Reduced Boltzmann Transport Equation (RBTE). With that in mind, we name the left side of the RBTE as $b^L$ and the right side as $b^R$. Notice that we can rewrite both sides of Eq. \eqref{reduced_BTE} as:
\begin{align}
\notag
    b^L &= v_\parallel \frac{\partial f_e^{\kappa\delta}}{\partial z} \ , \\
    &= v_\parallel \left[ \frac{\partial f_e^{\kappa\delta}}{\partial n_e} \frac{\partial n_e}{\partial z} + \frac{\partial f_e^{\kappa\delta}}{\partial W_{\parallel,e}}\frac{\partial W_{\parallel,e}}{\partial z} \right]  \ ,
    \label{b_L}
\end{align}
and
\begin{align}
    b^R = - \nu_e(\vec{v}) \left( f^{\kappa\delta}_e - f_e^\kappa \right) \,,
\end{align}
where we have considered that $f_e^{\kappa\delta}$ varies along the magnetic field direction $\hat z$.

We must keep in mind that $b^L = b^R$. 
To address the matter above, we consider minor asymmetries that manifest as small values of the parameter $\delta_e$. This enables us to employ a second-order Taylor expansion in the skewness parameter as a suitable approximation for the problem. This is,
\begin{align}
\notag
    f^{\kappa\delta}_e &\approx f_e^\kappa + \frac{\partial f^{\kappa\delta}_e}{\partial\delta_e} \bigg|_{\delta = 0} \delta_e + \frac{1}{2} \frac{\partial^2f^{\kappa\delta}_e}{\partial\delta_e^2} \bigg|_{\delta = 0} \delta_e^2 \ , \\
    &\approx f^{(0)}_e + f^{(1)}_e\delta_e + f^{(2)}_e\delta_e^2 \ .
    \label{f_approx}
\end{align}
similarly to the expansion done in \cite{Zenteno2021model}. Furthermore, it is imperative to acknowledge that, for macroscopic purposes, we intend to solve for the macroparameter $\delta_e$. Replacing Eq. \eqref{f_approx} in Eq. \eqref{reduced_BTE}, considering Eq. \eqref{b_L} now, we can integrate the RBTE across the entire velocity space. Therefore, we will obtain an equation $B^L = B^R$, where $B^S = \int b^S d^3v$ with $S = L, R$, namely:
\begin{align}
    B^L &= -\delta_eW_{\parallel,e} \frac{\partial n_e}{\partial z} \psi_1(\kappa) - \delta_en_{s} \frac{\partial W_{\parallel,e}}{\partial z} \psi_1(\kappa)  \ , \label{B_L} \\
    B^R &= \delta_e^2 n_e\nu_{e,0} \left[ \psi_2(\kappa) - \psi_3(\kappa) \right] \ .
    \label{B_R}
\end{align}
Here, $\{\psi(\kappa)_i\}_{i=1}^3$ are functions of $\kappa$ obtained from the derivation (see Appendix \ref{sec:appendix} for details). With Eqs. \eqref{B_L} and \eqref{B_R}, we can ultimately deduce a closed-form expression for the asymmetry parameter, thus arriving at the ensuing result:
\begin{equation}
    \delta_e = \Psi_\alpha(\kappa) \left( \frac{1}{\nu_{e,0}} \frac{\partial W_{\parallel,e}}{\partial z} + \frac{W_{\parallel,e}}{\nu_{e,0}n_e} \frac{\partial n_e}{\partial z} \right) \,,
    \label{delta_final}
\end{equation} 
where $\Psi_\alpha$ is a function of $\kappa$ as shown in Figure~\ref{Psi} (see Appendix \ref{sec:appendix}). 

Finally, considering that the mean free path can be expressed as $\lambda_{\rm{fp}} = W_{\parallel,e}/\nu_{e,0}$ we obtain
\begin{equation}
    \delta_e = \Psi_\alpha(\kappa)\,\left(\frac{1}{2}\,K_T + K_n\right)\,,
    \label{delta_knudsen}
\end{equation} 
where $K_T = \lambda_{\rm{fp}}/L_T$ and $K_n = \lambda_{\rm{fp}}/L_n$ are the temperature and density Knudsen numbers (here $L_T = T/|\nabla T|$, $L_n = n/|\nabla n|$ are the scale height of the temperature and density gradients). In the context of the solar wind, the mean free path $\lambda_{\rm fp}$ is directly related to the collisional age of the electrons, which measures the cumulative effect of Coulomb collisions during their propagation away from the Sun. Electrons with small collisional ages remain only weakly scattered, preserving signatures of non-equilibrium transport, while larger collisional ages correspond to more collisionally relaxed populations. The present formulation, based on Knudsen numbers constructed from $\lambda_{\rm fp}$ and macroscopic gradient scales, is therefore most naturally suited to describe regimes of small to moderate collisional age, where collisional effects are non-negligible but do not fully isotropize the distribution function. Through this procedure, we have successfully recovered reasonable and anticipated outcomes. Illustrative instances of these outcomes include the gradient-dependent terms, which pertain to both thermal velocity and number density.

Note that the relationship obtained in Equation \eqref{delta_knudsen} appears to be consistent with both the Spitzer \& Harm theory and the findings discussed in \cite{bale2013electron}. This can be primarily inferred from the linear relationship between the skewness parameter and the Knudsen number. Although the relationship presented in this study is between delta and $K_T$, it is expected that the heat flux also depends on the asymmetry of the distribution function as proposed by \cite{vinas2015electron,Zenteno2021model}, i.e., $\delta_e \sim q_e/q_{e,0} \sim K_T$ and observed by~\cite{eyelade2025characterizing}. However, in our study, we go a step further by additionally incorporating the effect of a potential density gradient, defining another characteristic spatial scale of the system. Thus, it becomes possible to define a Knudsen number corresponding to this scale, $K_n$. On the other hand, it is essential to emphasize that the scope of this result is linear as long as we consider $\delta_e$ to be sufficiently small. For larger values, higher-order terms in delta may need to be taken into account. While this calculation was not the primary focus of this research, it would be interesting to explore the nonlinear regime of our considerations and, if possible, investigate the saturation of the heat-flux curve as a function of the Knudsen number when the latter exceeds a certain value.

Beyond these theoretical considerations, recent observational analyses using the CS description on WIND/SWE-VEIS eVDFs have recovered the skewness of the solar wind electron distribution with a single asymmetry parameter $\delta_e$ and indicated that $\kappa$ is mainly independent of $\delta_e$ \citep{eyelade2025characterizing}. In our framework, this empirical decoupling is naturally accommodated because gradient-based Knudsen numbers primarily control the skewness parameter, while $\kappa$ only enters through the prefactor $\Psi_{\alpha}(\kappa)$, which rapidly saturates for large $\kappa$ (see Fig.~\ref{Psi}). Moreover, the scaling proposed here, $\delta_e$ increasing with $\tfrac{1}{2}K_T+K_n$, suggests concrete cross-checks with CS-derived heat-flux and $\delta_e$ across heliocentric distance, where variations in mean free path and macroscopic gradients are expected to imprint radial trends on $\delta_e$ and related transport proxies. In this sense, the CS model provides an immediate testbed for the gradient-controlled picture developed in this work, without presupposing any fixed relation between $\delta_e$ and $\kappa$. However, a fully systematic validation of this prediction would require independent estimates of the macroscopic gradients entering the Knudsen numbers. In practice, this remains challenging for in-situ spacecraft measurements, since reliable determinations of temperature and density scale lengths along the magnetic-field direction are difficult to obtain in the solar wind. Nevertheless, several observational strategies exist to approximate these gradients. For instance, studies of electron heat flux have already employed estimates of the temperature-gradient scale to construct Knudsen numbers in the solar wind \citep{bale2013electron}. In addition, empirical radial profiles derived from Helios electron observations provide estimates of large-scale temperature and density gradients in the expanding solar wind \citep{vstverak2015electron}. More recently, multi-spacecraft analyses combining data from missions such as Parker Solar Probe, Helios, and Ulysses have reconstructed radial trends of key plasma parameters across the heliosphere \citep{maruca2023trans}. Together with measurements of the skewness parameter $\delta_e$ from eVDF fits, these approaches suggest that the gradient-controlled scaling proposed here could be tested observationally by comparing $\delta_e$ with independently inferred effective Knudsen numbers.

\begin{figure}[ht]
\centering
    \includegraphics[width=0.8\textwidth]{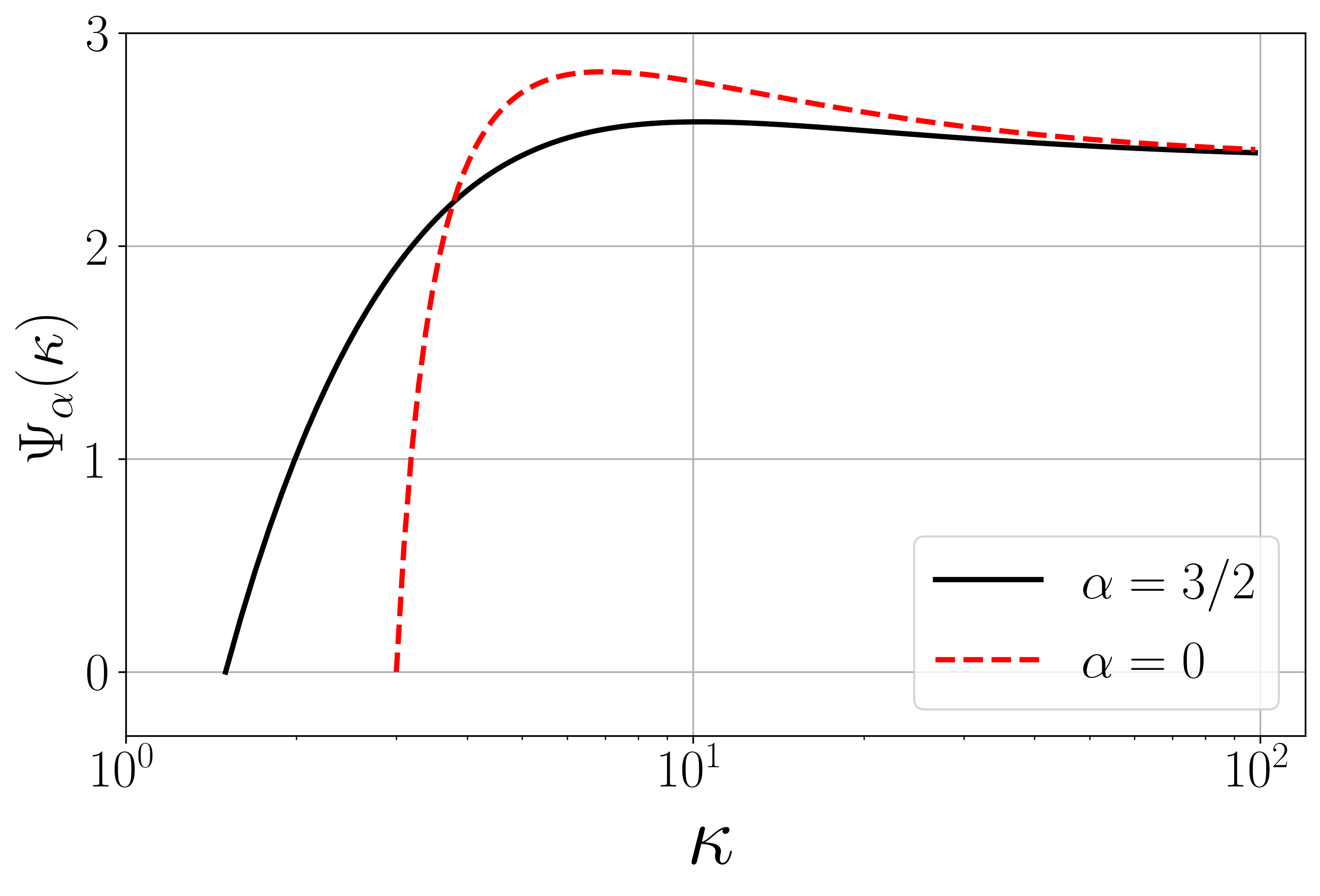}
    \caption{Comparative plot of $\Psi_\alpha$ factor, for two different values of $\alpha$. The dashed curve in red is for $\alpha = 0$, and the solid black line is for $\alpha = 3/2$. We can observe, independent of the value of $\alpha$, how the $\Psi_\alpha$ function saturates quickly for high values of $\kappa$.}
    \label{Psi}
\end{figure}

Regarding the theoretical foundations of our model, we note that the scope of our assumptions, inspired by Beck's work \citep{beck2000}, remains consistent throughout the presented study, irrespective of the particular ad hoc form adopted here for the collision frequency. This is particularly evident from Figure \ref{Psi}, where we observe that the $\Psi_\alpha(\kappa)$ factor, encompassing all information of the kappa parameter, converges to $\Psi_\alpha(\kappa) = 2.5$ for large values of kappa. This implies the initial assumption that delta does not depend directly on kappa is reasonably satisfied for kappa values greater than 4. Therefore, for most values of $\kappa$, there is no significant difference in the election of the type of Kappa distribution used to model the VDF of electrons. 

Furthermore, if Beck’s theory applies, a direct relationship can be established between the temperature and density gradients and the system's Reynolds number, as proposed in \citet{beck2000}. In this framework, the relevant control parameter is $c/\sqrt{R_\text{e}}$ (with $c$ being just a proportionality constant), which Beck argues must remain small in order to describe weak departures from equilibrium. Within our formulation, this requirement is fully
consistent with the small-skewness assumption ($\delta \ll 1$) adopted throughout this work. The corresponding relation can then be written as
\begin{equation}
       \frac{c}{\sqrt{R_\text{e}}} = \Psi_\alpha(\kappa)\,\left(\frac{1}{2}\,K_T + K_n\right)\, ,
    \label{K_Reynolds}
\end{equation}
which is therefore expected to hold in the regime where $c/\sqrt{R_\text{e}} \ll 1$. Since this quantity scales as $R_\text{e}^{-1/2}$, the condition naturally selects systems characterized by sufficiently large Reynolds numbers. In this sense, Eq.~(\ref{K_Reynolds}) should be regarded as a consistency relation linking the small-skewness limit to the turbulent properties of the plasma, rather than as an independent closure. The equation is valid if we treat the Reynolds number as a vorticity parameter. That is, as the Reynolds number increases, particles are expected to evolve in a flow increasingly dominated by turbulent fluctuations and coherent structures. From a transport perspective, these structures can enhance dispersion and effectively constrain the free flow of particles, reducing the effective mean free path relative to weakly turbulent conditions. Within this heuristic framework, an increase in $R_\text{e}$ can be associated with a decrease in the effective Knudsen number, in qualitative agreement with Equation~(\ref{K_Reynolds}).

The relation between skewness corrections and the Reynolds number has been experimentally investigated in laboratory turbulence. In particular, \citet{beck2001measuring} analyzed velocity increment distributions in a turbulent Taylor–Couette flow and showed that asymmetric corrections to the nonextensive probability density functions accurately reproduce the measured velocity distributions over a wide range of Reynolds numbers. In that experiment, the skewness term appears as a perturbative correction to the effective energy of the distribution, and its strength can be constrained directly from measured velocity statistics. These results provide an important experimental benchmark demonstrating that skewness corrections of the type considered here can indeed emerge in turbulent systems and that their magnitude may depend on macroscopic control parameters such as the Reynolds number. It is worth noting, however, that the Taylor–Couette experiment concerns probability distributions of velocity increments in a neutral fluid, whereas the present work focuses on velocity distribution functions in a plasma. Despite this difference, similar analyses based on velocity or magnetic-field increments in space plasma turbulence could provide a complementary avenue to test asymmetric corrections to non-Gaussian distributions in plasma environments \citep{pollock2018magnetospheric,chasapis2018situ,chhiber2020clustering}.

 Although this interpretation falls outside the main scope of the present work, it offers a plausible physical link among collisional transport, asymmetry, and turbulence that warrants further investigation.

\section{Conclusions}
\label{sec:conclusions}

Space plasmas, such as the solar wind, particularly the electron population, often exhibit velocity distribution functions that deviate from Maxwellian equilibrium, displaying both suprathermal tails and measurable asymmetries. In this work, we have addressed the origin of such asymmetries by considering a Skew--Kappa representation of the electron VDF and analyzing its stability within the framework of the Boltzmann Transport Equation. By incorporating a Krook-like collisional term, we have shown that the skewness parameter $\delta_e$ admits a closed-form expression and depends directly on macroscopic plasma quantities, such as the thermal velocity, density, and temperature gradients, as well as the effective collision frequency. This result places the asymmetry of the distribution function on the same footing as well-established transport parameters in space plasma physics, highlighting its relevance as a measurable non-thermal property of the system.

We emphasize that the results presented here rely on a well-defined set of assumptions that delimit the model's regime of validity. In particular, we have explicitly restricted our analysis to small values of the asymmetry parameter ($\delta_e \ll 1$) and assumed that its temporal evolution occurs on timescales much longer than the kinetic ones. This separation of scales allows the system to be described, at any given instant, by a quasi-stationary asymmetric equilibrium, effectively providing a snapshot of the electron VDF in a weakly evolving state. Within this framework, imposing a Skew--Kappa distribution as a solution of the Boltzmann transport equation introduces non-trivial constraints on the admissible form of the collision operator. In particular, the collision frequency cannot be chosen arbitrarily: simple models such as a constant collision frequency or prescriptions that diverge at $|\vec{v}| \to 0$ lead to inconsistent or trivial solutions of the reduced kinetic equation. Instead, a physically acceptable collision frequency must depend on velocity while remaining finite at low velocities. We have shown that a model satisfying these requirements yields a consistent and accurate approximation, enabling the derivation of a closed-form expression for the asymmetry parameter. Moreover, this approach naturally establishes connections between the VDF asymmetry and macroscopic plasma properties, recovering previously reported relationships between asymmetry, heat flux, and temperature gradients within a unified transport-based framework.

Consistent with this theoretical picture, recent Core--Strahlo fits to in-situ eVDFs recover the field-aligned asymmetry using a single parameter, $\delta_e$, and report only a weak empirical coupling with $\kappa$, providing observational support for the gradient-based interpretation proposed here \citep{eyelade2025characterizing}. Additionally, density gradients play a non-negligible role in controlling the distribution function's asymmetry. We further establish consistency with the framework proposed by \citet{beck2000}, showing that the skewness parameter exhibits only a weak dependence on the kappa index (see Fig.~\ref{Psi}), in agreement with the assumptions underlying that theory. In this context, the present results suggest that the skewness parameter $\delta_e$ may serve as an observational proxy of non-equilibrium transport in space plasmas, linking the velocity distribution function directly to macroscopic quantities such as temperature and density gradients through the effective Knudsen number. Although a fully systematic validation of this prediction requires reliable estimates of macroscopic gradients, previous observational studies have demonstrated that the ingredients entering the Knudsen number can be constrained in the solar wind. In particular, temperature-gradient scales have been inferred from electron heat-flux measurements \citep{bale2013electron}, while radial profiles derived from Helios observations and multi-spacecraft analyses provide estimates of large-scale gradients of plasma parameters across the heliosphere \citep{vstverak2015electron,maruca2023trans}. Combined with measurements of $\delta_e$ from eVDF fits, these approaches enable testing the proposed scaling by comparing the observed asymmetry with independently inferred effective Knudsen numbers.

An additional point worth highlighting is that our approach consolidates three perspectives that, although developed independently, converge into a consistent picture. On the one hand, \citet{vinas2015electron, Zenteno2021model} showed that under the Skew--Kappa approximation for solar wind electrons, the asymmetry of the velocity distribution function ($\delta_e$) scales linearly with the parallel electron heat flux, $q_{\parallel,e}$. On the other hand, the classical result of Spitzer \& Harm established that this heat flux is proportional to the Knudsen number $K_T$, as confirmed by \citet{bale2013electron}. Finally, in the present work, we have demonstrated that the skewness parameter itself is proportional to an effective Knudsen number, $K_N$, defined in terms of density and temperature gradients. Taken together, these three relationships form a closed logical framework that unifies skewness, heat transport, and collisionality through the Knudsen number, providing a coherent bridge between kinetic theory and observational proxies of space plasma dynamics.

\section*{Acknowledgments}

We are grateful for support from the postdoctoral project No. 0016500-2025 funded by the University of Chile, and from ANID, Chile, through a National Doctoral Scholarship No. 21182002 (IGM) and a FONDECYT grant No. 1240281 (PSM).  

\appendix

\section{Derivation Details}
\label{sec:appendix}
We will set the steady distribution as a Skew-Kappa distribution
\begin{align}
     f_e^{\kappa\delta} &= n_e A_{\kappa\delta}\left[   1 + \frac{v_\perp^2}{(\kappa - \alpha)W_{\perp,e}^2} + \frac{v_\parallel^2}{(\kappa - \alpha)W_{\parallel,e}^2} + \frac{{\delta_e}}{\kappa - \alpha}P\left( \frac{v_\parallel}{W_{\parallel,e}}\right) \right]^{-(\kappa + 1)} 
\end{align}
where $\delta_e$ is the skewness parameter, and $\alpha = 0, 3/2$. Furthermore, we define the polynomial function $P$ by $P(x) = x - x^3/3$.

We need to calculate the normalization factor, $A_{\kappa\delta}$. To accomplish this, let us establish the following:
\begin{align}
    g^{\kappa\delta}_e = \left[  1 + \frac{v_\perp^2}{(\kappa - \alpha)W_{\perp ,e}^2} + \frac{v_\parallel^2}{(\kappa - \alpha)W_{\parallel ,e}^2} + \frac{{\delta_e}}{\kappa - \alpha} P\left( \frac{v_\parallel}{W_{\parallel,e}}\right)  \right]^{-(\kappa + 1)} 
\end{align}
Given our initial premise that the distribution function exhibits subtlety in its asymmetry, i.e., small values of the skewness parameter, we expand in a second-order Taylor series in $\delta_e$, retaining only quadratic terms. Hence,
\begin{align}
\notag
    g^{\kappa\delta}_e &\approx g_e^\kappa + \frac{\partial g^{\kappa\delta}_e}{\partial\delta_e} \bigg|_{\delta = 0} \delta_e + \frac{1}{2} \frac{\partial^2g^{\kappa\delta}_e}{\partial\delta_e^2} \bigg|_{\delta = 0} \delta_e^2 \ , \\
    &\approx g^{(0)}_e + g^{(1)}_e\delta_e + g^{(2)}_e\delta_e^2 \ ,
    \label{g_approx}
\end{align}
where we have defined:
\begin{align}
    g^{(0)}_e &= g_e^\kappa \ , \\
    g^{(1)}_e &= - \frac{\kappa + 1}{\kappa - \alpha} P g^{\kappa+1}_e \ , \\
    g^{(2)}_e &= \frac{1}{2}\frac{(\kappa + 1)(\kappa + 2)}{(\kappa - \alpha)^2} P^2 g^{\kappa + 2}_e \ , 
\end{align}
and
\begin{align}
    g^{\kappa'}_e = \left[  1 + \frac{v_\perp^2}{(\kappa - \alpha)W_{\perp,e}^2} + \frac{v_\parallel^2}{(\kappa - \alpha)W_{\parallel,e}^2}  \right]^{-(\kappa' + 1)} \ .
\end{align}
With the foregoing, we can obtain the normalization factor, $A_{\kappa\delta}$. This is:
\begin{align}
\notag
    A_{\kappa\delta}^{-1} &= \int g^{\kappa\delta}_e \, d^3v \ , \\
    A_{\kappa\delta} &\approx A_\kappa \left( 1 - \frac{1}{4} \psi_0(\kappa) \delta_e^2 \right) \ ,
\end{align}
where,
\begin{align}
    A_\kappa = W_{\perp,e}^{-2}W_{\parallel,e}^{-1}\pi^{-3/2}(\kappa-\alpha)^{-3/2} \frac{\Gamma(\kappa+1)}{\Gamma(\kappa-1/2)} \ .
\end{align}
and
\begin{align}
    \psi_0(\kappa) = \frac{\kappa - 1/2}{\kappa - \alpha} + \frac{5}{12} \frac{\kappa - \alpha}{\kappa - 3/2} - 1 \ .
\end{align}
Let us consider equation \eqref{reduced_BTE} and take the difference between the left-hand and the right-hand side. We will refer to the left-hand side of the equation as:
\begin{align}
    b^L &= v_\parallel \frac{\partial f_e^{\kappa\delta}}{\partial z} \ , \\
    &= v_\parallel \left[ \frac{\partial f_e^{\kappa\delta}}{\partial n_e} \frac{\partial n_e}{\partial z} + \frac{\partial f_e^{\kappa\delta}}{\partial W_{\parallel,e}}\frac{\partial W_{\parallel,e}}{\partial z} \right]  \ .
\end{align}
On the other hand, the right-hand side of the equation would be
\begin{align}
    b^R = - \nu_e(\vec{v}) \left( f^{\kappa\delta}_e - f_e^\kappa \right) \ ,
\end{align}
Both sides of the equation must be considered and developed under the approximation shown in Eq. \eqref{g_approx}. Subsequently, since this treatment aims to obtain a statistical macro-parameter, $\delta_e$, we should integrate the above in velocity space. Let us say that $B^S = \int b^S d^3v$ with $S = L, R$. With this, we can obtain
\begin{align}
    B^L &= -\delta_eW_{\parallel,e} \frac{\partial n_e}{\partial z} \psi_1(\kappa) - \delta_en_{s} \frac{\partial W_{\parallel,e}}{\partial z} \psi_1(\kappa)  \ , 
\end{align}
and 
\begin{align}
    B^R = \delta_e^2 n_e\nu_{e,0} \left[ \psi_2(\kappa) - \psi_3(\kappa) \right] \ .
\end{align}
where $\psi_i$ are functions of $\kappa$ parameter, and these are defined by,
\begin{align}
    \psi_1(\kappa) &= \frac{1}{2}\left( 1 - \frac{1}{2}\frac{\kappa - \alpha}{\kappa - \frac{3}{2}} \right) \ , \\
    \psi_2(\kappa) &= \frac{1}{4}\frac{\psi_0(\kappa)\Gamma(\kappa + 1)^2}{\Gamma\left(\kappa - \frac{1}{2}\right)\Gamma\left(\kappa + \frac{5}{2}\right)} \ , \\
    \psi_3(\kappa) &= \frac{1}{4}\frac{(\kappa + 1)(\kappa + 2)}{\left(\kappa - \alpha\right)\left(\kappa - \frac{7}{2}\right)} \frac{\Gamma(\kappa)\Gamma(\kappa + 1)}{\Gamma\left(\kappa - \frac{1}{2}\right)\Gamma\left(\kappa + \frac{7}{2}\right)} \left[ (1 + \alpha)\kappa + \frac{5}{12}(\kappa - \alpha)^2 \right] \ .
\end{align}
Finally, solving for delta, we obtain the following:
\begin{align}
    \delta_e = \Psi_\alpha(\kappa) \left( \frac{1}{\nu_{e,0}} \frac{\partial W_{\parallel,e}}{\partial z} + \frac{W_{\parallel,e}}{\nu_{e,0}n_e} \frac{\partial n_e}{\partial z} \right) \,,
\end{align}
where $\Psi_\alpha(\kappa) = \psi_1(\kappa)/[\psi_3(\kappa) - \psi_2(\kappa)]$. 




\bibliography{biblio}{}

@article{pollock2018magnetospheric,
  title={Magnetospheric multiscale observations of turbulent magnetic and electron velocity fluctuations in Earth's magnetosheath downstream of a quasi-parallel bow shock},
  author={Pollock, Craig J and Burch, James L and Chasapis, Alexandros and Giles, Barbara L and Mackler, David A and Matthaeus, William H and Russell, Christopher T},
  journal={Journal of Atmospheric and Solar-Terrestrial Physics},
  volume={177},
  pages={84--91},
  year={2018},
  publisher={Elsevier}
}

@article{chhiber2020clustering,
  title={Clustering of intermittent magnetic and flow structures near Parker Solar Probe’s first perihelion—a partial-variance-of-increments analysis},
  author={Chhiber, Rohit and Goldstein, M L and Maruca, BA and Chasapis, A and Matthaeus, WH and Ruffolo, D and Bandyopadhyay, R and Parashar, TN and Qudsi, R and De Wit, T Dudok and others},
  journal={The Astrophysical Journal Supplement Series},
  volume={246},
  number={2},
  pages={31},
  year={2020},
  publisher={IOP Publishing}
}

@article{bruno2013solar,
  title={The solar wind as a turbulence laboratory},
  author={Bruno, Roberto and Carbone, Vincenzo},
  journal={Living Reviews in Solar Physics},
  volume={10},
  number={1},
  pages={1--208},
  year={2013},
  publisher={Springer}
}

@article{matthaeus2016ensemble,
  title={Ensemble space-time correlation of plasma turbulence in the solar wind},
  author={Matthaeus, WH and Weygand, JM and Dasso, S},
  journal={Physical Review Letters},
  volume={116},
  number={24},
  pages={245101},
  year={2016},
  publisher={APS}
}

@article{Jana2023Studies,
title={Studies of Transient Photoplasma Evolution in an Electrostatic Field: Single Particle Motion to Its Collective Behavior},
author={Jana, Biswajit and Dikshit, Biswaranjan},
journal={IEEE Transactions on Plasma Science},
year={2023},
volume={51},
pages={344-351},
doi={10.1109/TPS.2022.3233808}
}

@article{tsallis1988possible,
  title={Possible generalization of Boltzmann-Gibbs statistics},
  author={Tsallis, Constantino},
  journal={Journal of statistical physics},
  volume={52},
  number={1},
  pages={479--487},
  year={1988},
  publisher={Springer}
}

@article{vasyliunas1968survey,
  title={A survey of low-energy electrons in the evening sector of the magnetosphere with OGO 1 and OGO 3},
  author={Vasyliunas, Vytenis M},
  journal={Journal of Geophysical Research},
  volume={73},
  number={9},
  pages={2839--2884},
  year={1968},
  publisher={Wiley Online Library}
}

@article{hasegawa1985plasma,
  title={Plasma distribution function in a superthermal radiation field},
  author={Hasegawa, Akira and Mima, Kunioki and Duong-van, Minh},
  journal={Physical Review Letters},
  volume={54},
  number={24},
  pages={2608},
  year={1985},
  publisher={APS}
}

@article{pierrard2010kappa,
  title={Kappa distributions: Theory and applications in space plasmas},
  author={Pierrard, V and Lazar, M},
  journal={Solar physics},
  volume={267},
  number={1},
  pages={153--174},
  year={2010},
  publisher={Springer}
}

@article{lazar2015destabilizing,
  title={Destabilizing effects of the suprathermal populations in the solar wind},
  author={Lazar, Marian and Poedts, Stefaan and Fichtner, H},
  journal={Astronomy \& Astrophysics},
  volume={582},
  pages={A124},
  year={2015},
  publisher={EDP Sciences}
}

@article{lazar2017firehose,
  title={Firehose constraints of the bi-Kappa-distributed electrons: a zero-order approach for the suprathermal electrons in the solar wind},
  author={Lazar, Marian and Shaaban, SM and Poedts, Stefaan and {\v{S}}tver{\'a}k, {\v{S}}},
  journal={Monthly Notices of the Royal Astronomical Society},
  volume={464},
  number={1},
  pages={564--571},
  year={2017},
  publisher={Oxford University Press}
}

@book{lazar2021kappa,
  title={Kappa distributions},
  author={Lazar, Marian and Fichtner, Horst},
  year={2021},
  publisher={Springer}
}

@article{scherer2018regularized,
  title={Regularized $\kappa$-distributions with non-diverging moments},
  author={Scherer, Klaus and Fichtner, Horst and Lazar, Marian},
  journal={Europhysics Letters},
  volume={120},
  number={5},
  pages={50002},
  year={2018},
  publisher={IOP Publishing}
}

@article{husidic2020linear,
  title={Linear dispersion theory of parallel electromagnetic modes for regularized Kappa-distributions},
  author={Husidic, Edin and Lazar, Marian and Fichtner, Horst and Scherer, Klaus and Astfalk, Patrick},
  journal={Physics of Plasmas},
  volume={27},
  number={4},
  year={2020},
  publisher={AIP Publishing}
}

@article{beck2000,
  title={Application of generalized thermostatistics to fully developed turbulence},
  author={Beck, Christian},
  journal={Physica A: Statistical Mechanics and its Applications},
  volume={277},
  number={1-2},
  pages={115--123},
  year={2000},
  publisher={Elsevier}
}

@article{gallo2022langevin,
  title={Langevin based turbulence model and its relationship with Kappa distributions},
  author={Gallo-M{\'e}ndez, Iv{\'a}n and Moya, Pablo S},
  journal={Scientific Reports},
  volume={12},
  number={1},
  pages={1--8},
  year={2022},
  publisher={Nature Publishing Group}
}

@article{bhatnagar1954model,
  title={A model for collision processes in gases. I. Small amplitude processes in charged and neutral one-component systems},
  author={Bhatnagar, Prabhu Lal and Gross, Eugene P and Krook, Max},
  journal={Physical review},
  volume={94},
  number={3},
  pages={511},
  year={1954},
  publisher={APS}
}

@article{wang2017diffusion,
  title={The diffusion of charged particles in the weakly ionized plasma with power-law kappa-distributions},
  author={Wang, Lan and Du, Jiulin},
  journal={Physics of Plasmas},
  volume={24},
  number={10},
  pages={102305},
  year={2017},
  publisher={AIP Publishing LLC}
}

@article{wang2018viscosity,
  title={The viscosity of charged particles in the weakly ionized plasma with power-law distributions},
  author={Wang, Yue and Du, Jiulin},
  journal={Physics of Plasmas},
  volume={25},
  number={6},
  pages={062309},
  year={2018},
  publisher={AIP Publishing LLC}
}

@article{husidic2022toward,
  title={Toward a Realistic Evaluation of Transport Coefficients in Non-equilibrium Space Plasmas},
  author={Husidic, Edin and Scherer, Klaus and Lazar, Marian and Fichtner, Horst and Poedts, Stefaan},
  journal={The Astrophysical Journal},
  volume={927},
  number={2},
  pages={159},
  year={2022},
  publisher={IOP Publishing}
}

@article{shaaban2021advanced,
  title={Advanced interpretation of waves and instabilities in space plasmas},
  author={Shaaban, Shaaban M and Lazar, Marian and L{\'o}pez, Rodrigo A and Yoon, Peter H and Poedts, Stefaan},
  journal={Kappa distributions: From observational evidences via controversial predictions to a consistent theory of nonequilibrium plasmas},
  pages={185--218},
  year={2021},
  publisher={Springer}
}

@article{pilipp1987characteristics,
  title={Characteristics of electron velocity distribution functions in the solar wind derived from the Helios plasma experiment},
  author={Pilipp, WG and Miggenrieder, H and Montgomery, MD and M{\"u}hlh{\"a}user, K-H and Rosenbauer, H and Schwenn, R},
  journal={Journal of Geophysical Research: Space Physics},
  volume={92},
  number={A2},
  pages={1075--1092},
  year={1987},
  publisher={Wiley Online Library}
}

@article{beck2001measuring,
  title={Measuring nonextensitivity parameters in a turbulent Couette-Taylor flow},
  author={Beck, Christian and Lewis, Gregory S and Swinney, Harry L},
  journal={Physical Review E},
  volume={63},
  number={3},
  pages={035303},
  year={2001},
  publisher={APS}
}

@article{nieves2008solar,
  title={Solar wind electron distribution functions inside magnetic clouds},
  author={Nieves-Chinchilla, Teresa and Vi{\~n}as, Adolfo F},
  journal={Journal of Geophysical Research: Space Physics},
  volume={113},
  number={A2},
  year={2008},
  publisher={Wiley Online Library}
}

@inproceedings{vinas2015electron,
  title={Electron instability thresholds of solar wind magnetic fluctuations in non-thermal anisotropic kappa distribution plasmas: Survey of Wind-SWE-VEIS observations},
  author={Viñas, Adolfo F and Adrian, Mark L and Moya, Pablo S and Wendel, Deirdre E},
  booktitle={AGU Fall Meeting Abstracts},
  volume={2015},
  pages={SH33C--03},
  year={2015}
}

@article{Zenteno2023interplay,
year = {2023},
volume = {954},
pages = {184},
author = {Bea Zenteno-Quinteros and Pablo S. Moya and Marian Lazar and Adolfo F. Vi{\~n}as and Stefaan Poedts},
title = {Interplay between Anisotropy- and Skewness-driven Whistler Instabilities in the Solar Wind under the Core–Strahlo Model},
journal = {The Astrophysical Journal}
}

@article{zenteno2022role,
AUTHOR={Zenteno-Quinteros, Bea and Moya, Pablo S.},   
TITLE={The Role of Core and Strahlo Electrons Properties on the Whistler Heat-Flux Instability Thresholds in the Solar Wind},      
JOURNAL={Frontiers in Physics},      
VOLUME={10},           
YEAR={2022},      
DOI={10.3389/fphy.2022.910193}
}

@article{Zenteno2021model,
doi = {10.3847/1538-4357/ac2f9c},
year = {2021},
volume = {923},
number = {2},
pages = {180},
author = {Bea Zenteno-Quinteros and Adolfo F. Viñas and Pablo S. Moya},
title = {Skew-kappa Distribution Functions and Whistler Heat Flux Instability in the Solar Wind: The Core-strahlo Model},
journal = {The Astrophysical Journal}
}

@article{gallo2023understanding,
  title={Understanding the level of Turbulence by Asymmetric Distributions: a motivation for measurements in Space Plasmas},
  author={Gallo-M{\'e}ndez, Iv{\'a}n and Moya, Pablo S},
  journal={The Astrophysical Journal},
  volume={952},
  number={1},
  pages={30},
  year={2023},
  publisher={IOP Publishing}
}

@article{salem2003electron,
  title={Electron properties and Coulomb collisions in the solar wind at 1 AU: Wind observations},
  author={Salem, C and Hubert, Daniel and Lacombe, Catherine and Bale, Stuart D and Mangeney, Andr{\'e} and Larson, Davin E and Lin, RP},
  journal={The Astrophysical Journal},
  volume={585},
  number={2},
  pages={1147},
  year={2003},
  publisher={IOP Publishing}
}

@book{helander2005collisional,
  title={Collisional transport in magnetized plasmas},
  author={Helander, Per and Sigmar, Dieter J},
  volume={4},
  year={2005},
  publisher={Cambridge university press}
}

@article{bale2013electron,
  title={Electron heat conduction in the solar wind: transition from Spitzer--H{\"a}rm to the collisionless limit},
  author={Bale, SD and Pulupa, M and Salem, C and Chen, CHK and Quataert, E},
  journal={The Astrophysical Journal Letters},
  volume={769},
  number={2},
  pages={L22},
  year={2013},
  publisher={IOP Publishing}
}

@article{vstverak2015electron,
  title={Electron energetics in the expanding solar wind via Helios observations},
  author={{\v{S}}tver{\'a}k, {\v{S}}t{\v{e}}p{\'a}n and Tr{\'a}vn{\'\i}{\v{c}}ek, Pavel M and Hellinger, Petr},
  journal={Journal of Geophysical Research: Space Physics},
  volume={120},
  number={10},
  pages={8177--8193},
  year={2015},
  publisher={Wiley Online Library}
}

@article{maruca2023trans,
  title={The Trans-Heliospheric Survey-Radial trends in plasma parameters across the heliosphere},
  author={Maruca, Bennett A and Qudsi, Ramiz A and Alterman, BL and Walsh, Brian M and Korreck, Kelly E and Verscharen, Daniel and Bandyopadhyay, Riddhi and Chhiber, Rohit and Chasapis, Alexandros and Parashar, Tulasi N and others},
  journal={Astronomy \& Astrophysics},
  volume={675},
  pages={A196},
  year={2023},
  publisher={EDP Sciences}
}

@article{rosenbauer1977survey,
  title={A survey on initial results of the Helios plasma experiment},
  author={Rosenbauer, H and Schwenn, R and Marsch, E and Meyer, B and Miggenrieder, H and Montgomery, MD and Muehlhaeuser, KH and Pilipp, W and Voges, W and Zink, SM},
  journal={Journal of Geophysics Zeitschrift Geophysik},
  volume={42},
  number={6},
  pages={561--580},
  year={1977}
}

@article{feldman1978characteristic,
  title={Characteristic electron variations across simple high-speed solar wind streams},
  author={Feldman, WC and Asbridge, JR and Bame, SJ and Gosling, JT and Lemons, DS},
  journal={Journal of Geophysical Research: Space Physics},
  volume={83},
  number={A11},
  pages={5285--5295},
  year={1978},
  publisher={Wiley Online Library}
}

@article{lin1981energetic,
  title={Energetic electrons and plasma waves associated with a solar type III radio burst},
  author={Lin, RP and Potter, DW and Gurnett, DA and Scarf, FL},
  journal={Astrophysical Journal, Part 1, vol. 251, Dec. 1, 1981, p. 364-373.},
  volume={251},
  pages={364--373},
  year={1981}
}

@article{ogilvie2000electrons,
  title={Electrons in the low-density solar wind},
  author={Ogilvie, Keith W and Fitzenreiter, Richard and Desch, Michael},
  journal={Journal of Geophysical Research: Space Physics},
  volume={105},
  number={A12},
  pages={27277--27288},
  year={2000},
  publisher={Wiley Online Library}
}

@article{gosling2004dispersionless,
  title={Dispersionless modulations in low-energy solar electron bursts and discontinuous changes in the solar wind electron strahl},
  author={Gosling, JT and De Koning, CA and Skoug, RM and Steinberg, JT and McComas, DJ},
  journal={Journal of Geophysical Research: Space Physics},
  volume={109},
  number={A5},
  year={2004},
  publisher={Wiley Online Library}
}

@article{maksimovic2005radial,
  title={Radial evolution of the electron distribution functions in the fast solar wind between 0.3 and 1.5 AU},
  author={Maksimovic, Milan and Zouganelis, Ioannis and Chaufray, J-Y and Issautier, Karine and Scime, Earl E and Littleton, JE and Marsch, Eckart and McComas, DJ and Salem, C and Lin, RP and others},
  journal={Journal of Geophysical Research: Space Physics},
  volume={110},
  number={A9},
  year={2005},
  publisher={Wiley Online Library}
}

@article{pagel2007scattering,
  title={Scattering of suprathermal electrons in the solar wind: ACE observations},
  author={Pagel, Christina and Gary, S Peter and De Koning, Curt A and Skoug, Ruth M and Steinberg, John T},
  journal={Journal of Geophysical Research: Space Physics},
  volume={112},
  number={A4},
  year={2007},
  publisher={Wiley Online Library}
}

@article{vstverak2009radial,
  title={Radial evolution of nonthermal electron populations in the low-latitude solar wind: Helios, Cluster, and Ulysses Observations},
  author={{\v{S}}tver{\'a}k, {\v{S}}t{\v{e}}p{\'a}n and Maksimovic, Milan and Tr{\'a}vn{\'\i}{\v{c}}ek, Pavel M and Marsch, Eckart and Fazakerley, Andrew N and Scime, Earl E},
  journal={Journal of Geophysical Research: Space Physics},
  volume={114},
  number={A5},
  year={2009},
  publisher={Wiley Online Library}
}

@article{lopez2019particle,
  title={Particle-in-cell simulations of the whistler heat-flux instability in solar wind conditions},
  author={L{\'o}pez, RA and Shaaban, SM and Lazar, Marian and Poedts, Stefaan and Yoon, PH and Micera, Alfredo and Lapenta, Giovanni},
  journal={The Astrophysical Journal Letters},
  volume={882},
  number={1},
  pages={L8},
  year={2019},
  publisher={IOP Publishing}
}

@article{lopez2020alternative,
  title={Alternative high-plasma beta regimes of electron heat-flux instabilities in the solar wind},
  author={L{\'o}pez, RA and Lazar, M and Shaaban, SM and Poedts, Stefaan and Moya, PS},
  journal={The Astrophysical Journal Letters},
  volume={900},
  number={2},
  pages={L25},
  year={2020},
  publisher={IOP Publishing}
}

@article{bervcivc2020coronal,
  title={Coronal electron temperature inferred from the strahl electrons in the inner heliosphere: Parker solar probe and Helios observations},
  author={Ber{\v{c}}i{\v{c}}, Laura and Larson, Davin and Whittlesey, Phyllis and Maksimovi{\'c}, Milan and Badman, Samuel T and Landi, Simone and Matteini, Lorenzo and Bale, Stuart D and Bonnell, John W and Case, Anthony W and others},
  journal={The Astrophysical Journal},
  volume={892},
  number={2},
  pages={88},
  year={2020},
  publisher={IOP Publishing}
}

@article{halekas2020electrons,
  title={Electrons in the young solar wind: First results from the parker solar probe},
  author={Halekas, JS and Whittlesey, P and Larson, DE and McGinnis, D and Maksimovic, M and Berthomier, Matthieu and Kasper, JC and Case, AW and Korreck, KE and Stevens, ML and others},
  journal={The Astrophysical Journal Supplement Series},
  volume={246},
  number={2},
  pages={22},
  year={2020},
  publisher={IOP Publishing}
}

@article{halekas2021electron,
  title={Electron heat flux in the near-Sun environment},
  author={Halekas, Jasper S and Whittlesey, Phyllis L and Larson, Davin E and McGinnis, Daniel and Bale, Stuart D and Berthomier, Matthieu and Case, Anthony W and Chandran, Benjamin DG and Kasper, Justin Christophe and Klein, Kristopher G and others},
  journal={Astronomy \& Astrophysics},
  volume={650},
  pages={A15},
  year={2021},
  publisher={EDP Sciences}
}

@article{eyelade2025characterizing,
	author = {Eyelade, Adetayo V. and {Zenteno-Quinteros, Bea} and {Moya, Pablo S.} and {Silva, Javier I.} and {Urra, Benjamin A.} and {Lazar, Marian} and {Viñas, Adolfo F.}},
	title = {Characterizing solar wind electrons with the core-strahlo model: WIND-SWE-VEIS observations},
	DOI= "10.1051/0004-6361/202555368",
	url= "https://doi.org/10.1051/0004-6361/202555368",
	journal = {A\&A},
	year = 2025,
	volume = 702,
	pages = "A198",
}

@article{daniel2024magnetic,
  title={Magnetic spectra comparison for kappa-distributed whistler electron fluctuations},
  author={Daniel, HP and Moya, Pablo S and Zenteno-Quinteros, Bea and L{\'o}pez, Rodrigo A},
  journal={The Astrophysical Journal},
  volume={970},
  number={2},
  pages={132},
  year={2024},
  publisher={IOP Publishing}
}

@article{chasapis2018situ,
  title={In situ observation of intermittent dissipation at kinetic scales in the Earth's magnetosheath},
  author={Chasapis, Alexandros and Matthaeus, WH and Parashar, TN and Wan, M and Haggerty, CC and Pollock, CJ and Giles, BL and Paterson, WR and Dorelli, J and Gershman, DJ and others},
  journal={The Astrophysical Journal Letters},
  volume={856},
  number={1},
  pages={L19},
  year={2018},
  publisher={IOP Publishing}
}
\bibliographystyle{aasjournal}



\end{document}